\documentclass{ws-procs9x6}
\usepackage{graphicx}

\newcommand{\nn}{\nonumber\\}
\newcommand{\ph}{\varphi}
\newcommand{\ep}{\varepsilon}
\newcommand{\MSbar}{{\ensuremath{\overline{\mathrm{MS}}}}}
\renewcommand{\c}[1]{\mathcal{#1}}

\begin{document}

\title{Renormalization and resummation in field theories} 
\author{A. Jakov\'ac\footnote{E-mail: jakovac@planck.phy.bme.hu}}

\address{HAS Research Group ``Theory of Condensed Matter'' and TU
  Budapest, H-1521 Budapest, Hungary}

\author{Zs. Sz\'ep\footnote{E-mail: szepzs@achilles.elte.hu}}
\address{Research Group for Statistical Physics of the
  Hungarian Academy of Sciences, H-1117 Budapest, Hungary}

\maketitle

\abstracts{ Using resummation in perturbation theories at finite
  temperature or in non-equilibrium is unavoidable to obtain
  consistent results. Resummation, however, is often in conflict with
  renormalization. In this talk we give two possible solutions to
  problems of this type in static resummation. One is the use of
  auxiliary finite temperature UV divergent counterterms. The other is
  to use a renormalization scheme fitted to the environment. We also
  suggest a way how one could proceed in case of dynamic resummations.
}

\section{Introduction}
\label{sec:intro}

In perturbation theories at finite temperature and non-equilibrium
environment we are often faced with infrared (IR) divergent
diagrams. Although this divergency can be of physical origin, but more
often it is just an artefact of a not properly organized perturbation
series, and it disappears after summing up a certain subset of
diagrams. 

A zero temperature example is the on-shell singularity of the
perturbation series of a propagator. In this case one uses 1PI
resummation (Schwinger-Dyson equations) that makes possible to shift
the inverse propagator. At finite temperature a similar phenomenon
occurs since the environment modifies the properties of the
propagating particles. As a consequence we have to count with
temperature dependent mass terms\cite{DolanJackiw}, or
self-energies\cite{BrPis}. In these cases the Schwinger-Dyson
resummation is again necessary to avoid divergent perturbation series.
Coupling constant resummation is needed near a second order phase
transition point\cite{secondorder} or in case of non-Abelian gauge
theories\cite{BrPis}.

In many cases it happens, however, that new, unexpected, temperature
(or, more generally, environment) dependent ultraviolet (UV)
divergences appear. An illustrative example is the $\Phi^4$ model. Here
the one-loop correction to the self-energy at finite temperature and in
dimensional regularization reads as
\begin{equation}
  \raisebox{1.1em}{\emph{m}}\hspace*{-2.3em}
  \raisebox{-0.9em}{\includegraphics[height=1cm]{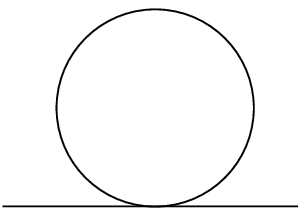}}\! =
  \displaystyle \frac{m^2}{16\pi^2} \left[-\frac1\ep +\gamma_E-1
    +\ln\frac{m^2}{4\pi\mu^2} \right] +
  \frac1{2\pi^2}\int\limits_m^\infty \!d\omega\,\sqrt{\omega^2-m^2}
  \,n(\omega).
\end{equation}
This diagram is UV divergent, so we need a mass counterterm $\delta
m^2 = \frac{m^2}{16\pi^2}\;\frac1\ep$. If we resum the mass we
effectively substitute $m^2\to M_T^2$, a temperature-dependent mass
term, and the same diagram now requires $\delta m^2 =
\frac{M_T^2}{16\pi^2}\;\frac1\ep$. This is $T$ and $\ep$ dependent in
the same time, in contradiction with the physical picture that
short distance physics cannot be affected by the physics in the IR.

In order to understand the problem deeper let us try to understand,
why, in general, resummation and renormalization can be in conflict.
The theory that defines the system is, in fact, the bare theory. The
renormalized theory is already a resummation in the bare theory: the
bare parameters are divided into sum of infinite parts (counterterms)
and these parts are taken into account at different levels of
perturbation theory. In this subdivision the organizing principle is
the UV relevance of the diagrams. There is a subtle ballance between
the diagrams and counterterms, and a series of theorems
ensure\cite{Collins} that this method finally works consistently for a
certain set of theories (the perturbatively renormalizable ones).

If we encounter other type of divergencies at finite energy scales we
are forced to do another resummation. In the course of this IR
resummation we order the diagrams according to their IR relevance.
``Normal'' and counterterm diagrams, however, have different IR
relevance in general, and so IR resummation will segregate them. So UV
divergences do not drop out, the resummed theory will be UV
inconsistent.

The above analysis suggest immediately a way out of this problem: we
should resum the appropriate counterterm diagrams together with the
most IR relevant ``normal'' diagrams\cite{JakSzep}. Here we describe
how to do that.

\section{Counterterm resummation by hand}
\label{sec:compcount}

We demonstrate the method for the mass resummation, but it will be
applicable for other cases (coupling constant resummations) as well.
We start from the generic procedure of resummation: we add to and
subtract from the Lagrangian the same term and treat the two parts in
different order in perturbation theory:
\begin{equation}
  -\c L_\mathrm{mass} =\underbrace{\frac{m^2+\Delta M^2}2\ph^2
        \rule[-1.3em]{0em}{3em}}_{\mbox{\small tree level}} +
        \underbrace{\left( \frac{\delta m^2}2\ph^2 - \frac{\Delta
        M^2}2\ph^2\right)}_{\mbox{\small one loop}},
\end{equation}
where the subscripts under the brace denote the order where the given
term starts to contribute. If $m^2=m_\mathrm{bare}^2,\,\delta
m_\mathrm{bare}^2 =0$ and $m_\mathrm{bare}^2+\Delta M^2=m_R^2$, then
we describe renormalization. If $\Delta M^2$ is temperature dependent,
then we have the ``thermal counterterm'' method\cite{thermalmass}. In
this case, however, as we have seen, $\delta m^2$ must be temperature
dependent to make the theory finite at one loop level.

To resum counterterms we can follow the same strategy, \emph{but now
  at one loop level}. We write
\begin{equation}
  -\c L_\mathrm{mass} =\underbrace{\frac{m_R^2+\Delta M^2}2\ph^2
        \rule[-1.3em]{0em}{3em}}_{\mbox{\small tree level}} +
        \underbrace{\left( \frac{\delta m^2+\delta m_T^2}2\ph^2 - \frac{\Delta
        M^2}2\ph^2\right)}_{\mbox{\small one loop}} +
        \underbrace{\left( -\frac{\delta
        m_T^2}2\ph^2\right)}_{\mbox{\small two loop}}.
\end{equation}
$\delta m_T^2$, referred as ``compensating counterterm'' should be
temperature dependent \emph{and} divergent (in $\Phi^4$ model $\delta
m_T^2=\frac{\Delta M^2}{16\pi^2}\;\frac1\ep$); but its value is
irrelevant from the point of view of the complete theory, being
subtracted one loop later. Otherwise it is a normal counterterm, its
value should be determined order by order. In mass-independent schemes
(where $\delta m^2 = z m^2$, where $z$ is mass-independent), however,
we expect that the shift of the tree level mass $m^2\to m^2+\Delta
M^2$ will imply the mass counterterm shift $\delta m^2 = zm^2 \to
zm^2+z\Delta M^2$, so we get $\delta m_T^2=z\Delta M^2$. This is not
the most general case (any finite term could be added), but an
especially comfortable one. In this case, namely, we arrive at
\begin{equation}
  \Pi^\mathrm{resum,\ \MSbar}(m^2+\Delta M^2) = \Pi^\MSbar(m^2)
  \biggr|_{m^2\to m^2+\Delta M^2},
\end{equation}
ie. we do the \emph{renormalization first}, and then substitute the
resummed mass into the finite expression. It can be
proven\cite{JakSzep} that this choice is in fact consistent to all
orders in perturbation theory.

\section{Resummation scheme}
\label{sec:ressch}

We can make the procedure of the previous subsection more systematic
by noting that the value of the tree level mass in renormalized
perturbation theory is not fixed, it is in fact the consequence of the
scheme chosen. The previous subsection was, from this point of view,
the description of a scheme where the tree level mass is
$m_\MSbar^2+\Delta M^2$.  We will call this scheme \emph{resummation
  scheme}. In a more generic framework resummation scheme is a scheme
that suits well the perturbation theory in an external
environment\cite{envfriend}. We define it that the infinite pieces of
the counterterms cancel the UV divergences, while the finite pieces
should diminish IR sensitivity of the theory.

As this definition shows, counterterms in a resummation scheme will
depend on the environment. Therefore changing the environment will
imply a change in the scheme; but results in different schemes are not
comparable\footnote{For example the mass in the finite temperature on
  shell scheme is always $m^2$: there is no explicit dependence on the
  temperature!}. We should project the results onto a fixed reference
scheme (we chose \MSbar\ for concreteness). Since we change from one
scheme to another, the projections can be done by adapting the
renormalized parameters of the Lagrangian\cite{Collins}. In case of
$\Phi^4$ theory there exist functions
\begin{equation}
  \label{eq:functs}
  m=m(m_\MSbar,\lambda_\MSbar,T),\quad \lambda = \lambda(m_\MSbar,
  \lambda_\MSbar, T),\quad \zeta=\zeta(m_\MSbar,\lambda_\MSbar,T),
\end{equation}
that for any $n$-point function
\begin{equation}
  G^{(n)}_\MSbar(p_i;m_\MSbar,\lambda_\MSbar) = \zeta^n
  G^{(n)}_\mathrm{resum}(p_i;m,\lambda).
\end{equation}
In perturbation theory this equation holds up to a given order: the
difference is the effect of the resummation performed by the
resummation scheme.

Technically we determine \eqref{eq:functs} from the condition that the
bare parameters should be the same\cite{Collins}: $m_\MSbar^2+\delta
m_\MSbar^2 = m^2 + \delta m^2$ in case of mass resummation. In the
on-shell scheme we obtain from this condition
\begin{equation}
  m^2 = m_\MSbar^2 + \Pi^\MSbar(m^2).
\end{equation}
This matching equation is just the gap equation in the language of the
resummation. 

At one loop level an equivalent form could have been reached if we had
wrote $m_\MSbar^2$ as the argument of $\Pi^\MSbar$. The two cases
correspond to different resummations: one with $\Pi^\MSbar(m^2)$
yields the super-daisy, while the other with $\Pi^\MSbar(m_\MSbar^2)$
the daisy resummation.

Resummation scheme can be used to resum any parameters of the
Lagrangian in a momentum-independent way (or at a fixed momentum). For
example we can resum the coupling constant in $\Phi^4$ theory. The
details are in Ref.\cite{JakSzep}. The results for the matching
equations \eqref{eq:functs} read in high temperature expansion
\begin{eqnarray}
  &&\frac1\lambda -\frac1{\lambda_\MSbar} =  \frac T{16\pi M} -
  \frac3{32\pi^2} \ln\frac{T^2}{e\tilde\mu^2}\nn
  &&m^2 -m^2_\MSbar = \frac{\lambda_\MSbar T^2}{24} -
  \frac{\lambda_\MSbar T m}{8\pi} +\frac{\lambda_\MSbar m^2}{32\pi^2}
  \ln\frac{T^2}{\tilde\mu^2}.
\end{eqnarray}
At high temperature $m^2 = m^2_\MSbar(T) + \frac1{24}\lambda_\MSbar(T)
T^2$, where the arguments refer to the renormalization scale in
\MSbar\ scheme. Near the critical point $m\sim T-T_c$, signaling a
second order phase transition. The coupling constant at high
temperatures behaves as $\lambda\sim\sqrt{\lambda_\MSbar(T)}$,
characteristic to a 3D, dimensionally reduced theory. Near the
critical point $\lambda\sim m\to 0$.

\section{Outlook}

So far we have spoken about momentum-independent resummations. The
whole procedure goes through, however, if momentum dependence is
concentrated only to the IR regime: for example when we have
$\lambda(p\to\infty)=\lambda_\mathrm{resum}$. In fact the UV behavior
is much more under control, and it is not affected by the IR problems
of the theory. So a possibility for momentum dependent cases would be
to make a partial resummation of the IR momentum dependence, and leave
the UV part for perturbation theory. This work is in progress.

\end{document}